\documentstyle[epsfig,aps,pra]{revtex}

\begin{document}
\title{Revivals of Coherence in Chaotic Atom-Optics Billiards}
\author{M. F. Andersen, T. Gr\"{u}nzweig, A. Kaplan and N. Davidson}
\address{Department of Physics of Complex Systems, Weizmann Institute of Science,\\
Rehovot 76100, Israel}
\date{\today}
\maketitle

\begin{abstract}
We investigate the coherence properties of thermal atoms confined
in optical dipole traps where the underlying classical dynamics is
chaotic. A perturbative expression derived for the coherence of
the echo scheme of [Andersen et. al., Phys. Rev. Lett. {\bf 90},
023001 (2003)] shows it is a function of the survival probability
or fidelity of eigenstates of the motion of the atoms in the trap.
The echo coherence and the survival probability display ''system
specific'' features, even when the underlying classical dynamics
is chaotic. In particular, partial revivals in the echo signal and
the survival probability are found for a small shift of the
potential. Next, a "semi-classical" expression for the averaged
echo signal is presented and used to calculate the echo signal for
atoms in a light sheet wedge billiard. Revivals in the echo
coherence are found in this system, indicating they may be a
generic feature of dipole traps.
\end{abstract}




\section{Introduction}

The capability to control and manipulate quantum systems that has
been developed in recent years holds promise for applying quantum
systems to information processing \cite{Nielsen00}. This possible
application has drawn much attention to the subject of
decoherence/dephasing, where the concept of ''fidelity'' and
''survival probability'' plays a major role. Some of the possible
quantum information schemes involve superposition states of the
internal states of trapped neutral atoms or ions. Interaction with
the trapping potential often causes these superposition states to
dephase, but, in analogy with the spin echo \cite{Hahn50}, a
coherence "echo" can be used to preserve the coherence
\cite{rowe02} \cite{andersen03} \cite{kuhr03}. Under certain
conditions the echo coherence of trapped atoms is limited by the
quantum dynamics of the atoms in the trap, thereby enabling study
of quantum dynamics by echo techniques, but also stressing that
the quantum dynamics of the atoms in the trap is of great
importance to the coherence of their internal states
\cite{andersen03}.

In this paper the dephasing of trapped atoms in coherent
superposition states of their internal states is investigated. A
special focus is put on neutral atoms trapped in optical dipole
traps, where the underlying classical dynamics is chaotic namely
atom-optics billiards \cite{milner01}\cite{Friedman01}. This is
done partly because chaotic billiards where used in the past as a
simplified model to describe decoherence \cite{jalabert01}, and
partly because they can have very long coherence times
\cite{andersen03b}. The atoms are considered ''hot'', meaning that
their temperature is much larger than the mean level spacing of
the trap, so they occupy high excited states in the trap, and an
output of the experiment is an ensemble average over many occupied
states.

The strength of the dipole potential depends on the internal state
of the atom. This causes atoms initially in coherent superposition
states of their internal degree of freedom but with different
external initial conditions to acquire different phases thereby
loosing the macroscopic coherence. This dephasing can be
suppressed by echo techniques, where each part of the
superposition state spends an equal amount of time in each
potential, thereby ensuring that the phases acquired are equal
\cite{andersen03}\cite{kuhr03}. However, since the external
potentials are slightly different (we denote this difference "the
perturbation"), also the dynamics depends on the internal state,
and loss of coherence from this is not cancelled by echo
techniques. For far-off -resonance optical dipole traps such a
perturbation is small, so a perturbative result is of great
interest. In section \ref{per} a perturbative expression for the
echo signal is derived, and confirmed numerically, showing that
the echo signal can be written as a function of the survival
probability or fidelity of eigenstates of the motion of the atoms
in the trap. Based on this perturbative result the long-time echo
coherence of atomic ensembles is predicted.

In the perturbative limit the echo coherence and the survival
probability display ''system specific'' features, despite the fact
that the underlying classical dynamics is ergodic\cite{Cohen01}.
This is demonstrated in section \ref{fire} by showing partial
revivals in the echo signal and the survival probability for a
perturbation consisting of a small spacial shift of the potential.
In section \ref{fem} a "semiclassical" perturbative expression for
the averaged echo coherence is derived. Two billiard systems are
considered throughout the paper. First, an "idealized" system - a
shift perturbation of the hard-wall Bunimovich billiard
\cite{burni79} where eigenstates can be computed efficiently
\cite{vergini95} is used to confirm the validity of the
perturbative and semiclassical approximations by comparing them to
a full quantum calculation (sections \ref{fire} and \ref{fem}).
These approximations are then used in section \ref{seks} to
calculate the echo coherence for micro wave (MW) spectroscopy of
atoms in an experimental realizable atom optics billiard, namely
the light-sheet wedge billiard \cite{milner01} \cite{Davidson95},
where revivals in analogy with those in the hard-wall stadium are
found. This suggest that revivals of coherence may be a generic
feature of optical dipole traps.

\section{Micro Wave Spectroscopy and Quantum Dynamics of trapped Atoms}

In this section pulsed micro wave spectroscopy on optical trapped
alkali atoms is considered.

\begin{figure}[b]
\includegraphics[width=3inc]{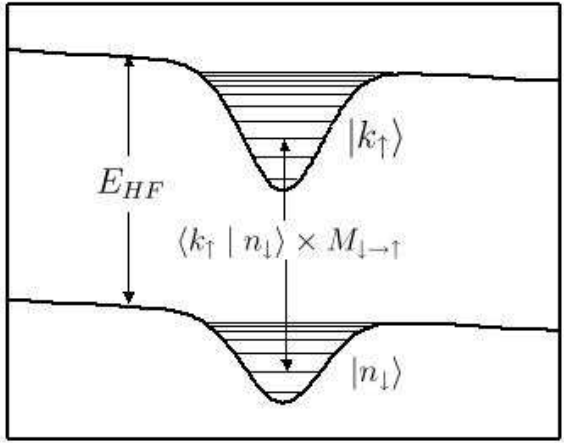}
\caption{The matrix elements for MW-transitions depend on the
overlap of the external motion states. The trap shown is a
Gaussian trap, and gravity gives the slope.} \label{fi1}
\end{figure}

The ground state of alkali atoms contains two hyperfine states
with an energy splitting of $E_{HF}=\hbar \omega _{HF}$, each
containing several magnetic substates. The hyperfine energy
splitting in Caesium atoms is the foundation of modern time
scales, so MW spectroscopy of this transition in an isolated
environment is of great importance and is therefore a very
advanced field. However, many quantum information processing
schemes require trapped atoms or ions, and if neutral alkali atoms
are placed inside an optical trapping potential, then the energy
levels are changed. Since the optical dipole potential is
inversely proportional to the detuning of the trap laser from
resonance \cite {CCT91} and the detuning will differ by the
hyperfine splitting, then there is a slight difference in
potential for atoms in different hyperfine states, even if the
states considered have the same matrix element for the dipole
interaction with the trapping light. For such two states (denoted
$\left| \uparrow \right\rangle $ and $\left| \downarrow
\right\rangle $) the Hamiltonian considered can be written as,
\begin{equation}
H=H_{\downarrow }\left| \downarrow \right\rangle \left\langle
\downarrow \right| + \left( H_{\uparrow }+E_{HF} \right) \left|
\uparrow \right\rangle \left\langle \uparrow \right| =\left(
\frac{p^{2}}{2m}+V_{\downarrow }\left( {\bf x}\right) \right)
\left| \downarrow \right\rangle \left\langle \downarrow \right|
+\left( \frac{p^{2}}{2m}+V_{\uparrow }\left( {\bf x}\right)
+E_{HF}\right) \left| \uparrow \right\rangle \left\langle \uparrow
\right| , \label{ham}
\end{equation}
where $p$ is the atomic center of mass momentum and $V_{\downarrow }$ [$%
V_{\uparrow }$] the external potential for an atom in state
$\left| \downarrow \right\rangle $ [$\left| \uparrow \right\rangle
$]. Typically $V$ is much smaller than $E_{HF}$ and $\left(
V_{\uparrow} - V_{\downarrow} \right)/V_{\downarrow} \sim
10^{-3}-10^{-5}$ \cite{andersen03}\cite{kuhr03}. The eigenenergies
of this Hamiltonian consist of two manifolds (belonging to $\left|
\downarrow \right\rangle $ and $\left| \uparrow \right\rangle $)
separated in energy by $E_{HF}$ (see Fig. \ref{fi1}).

The initial step in MW spectroscopy is to prepare all the
atoms in one of the spin states involved ($\left| \downarrow \right\rangle $%
), so their total wave function can be written as $\left|
\downarrow \right\rangle \otimes \left| \psi _{i}\right\rangle
\equiv \left| \downarrow ,\psi _{i}\right\rangle $, where $\left|
\psi _{i}\right\rangle $ represents the motional (external degree
of freedom) part of their wave function. If atoms prepared in
$\left| \downarrow ,\psi _{i}\right\rangle $ are irradiated with a
MW field close to resonance with $\omega _{HF}$, then transitions
between eigenstates belonging to different manifolds can be
driven. The matrix elements for these transitions is given by
$C_{nk}=\left\langle k_{\uparrow }\mid n_{\downarrow
}\right\rangle \times M_{\downarrow \rightarrow \uparrow }$ where
$M_{\downarrow \rightarrow \uparrow }$ is the
free space matrix element for the internal state transition, and $%
\left\langle k_{\uparrow }\mid n_{\downarrow }\right\rangle $ is
the overlap between $\left| n_{\downarrow }\right\rangle $, the
initial eigenstate of $H_{\downarrow }$, and $\left| k_{\uparrow
}\right\rangle $, the final eigenstate of $H_{\uparrow }$ (here
the very small momentum of the MW photon is neglected). Since in
general each of the eigenstates in the $\left| \downarrow
\right\rangle $-manifold is coupled to many eigenstates in the $
\left| \uparrow \right\rangle $-manifold by the MW-field, there is
no simple prediction of the wave function after MW irradiation.
However, in the simple limit of very short MW-pulses the external
wave function is left unchanged and only the internal state of the
atom is changed. Such a short pulse that puts atoms initially in
$\left| \downarrow ,\psi _{i}\right\rangle $, into the coherent
superposition state $\frac{1}{\sqrt{2}}\left| \downarrow ,\psi
_{i}\right\rangle +\frac{i}{\sqrt{2}}\left| \uparrow ,\psi
_{i}\right\rangle
$ is called a $\frac{\pi }{2}$-pulse and a pulse that flips the spin to $%
i\left| \uparrow ,\psi _{i}\right\rangle $ is called a $\pi $-pulse.

In the widely used Ramsey spectroscopy technique \cite{Ramsey56},
atoms initially prepared in $\left| \downarrow ,\psi
_{i}\right\rangle $ (assumed to be a pure state) and irradiated
with a $\frac{\pi }{2}$-pulse to generate the wave function
$\frac{1}{\sqrt{2}}\left| \downarrow ,\psi _{i}\right\rangle
+\frac{i}{\sqrt{2}}\left| \uparrow ,\psi _{i}\right\rangle $.
After some time $\tau $ this state will, in the rotating frame of
the MW
field, evolve into $\frac{1}{\sqrt{2}}\exp \left( -i\frac{H_{\downarrow }}{%
\hbar }\tau \right) \left| \downarrow ,\psi _{i}\right\rangle +\frac{i}{%
\sqrt{2}}\exp \left( -i\left( \frac{H_{\uparrow }}{\hbar }+\Delta
\right) \tau \right) \left| \uparrow ,\psi _{i}\right\rangle $,
where $\Delta = \omega_{HF} - \omega_{MW}$ is the MW-detuning.
Then the atoms are irradiated with a second $\frac{\pi }{2}$-pulse
generating the wave function $\frac{1}{2} \left( \exp \left(
-i\frac{H_{\downarrow }}{\hbar }\tau \right) -\exp \left( -i\left(
\frac{H_{\uparrow }}{\hbar }+\Delta \right) \tau \right) \right)
\left| \downarrow ,\psi _{i}\right\rangle +\frac{i}{2}\left( \exp
\left( -i\left( \frac{H_{\uparrow }}{\hbar }+\Delta \right) \tau
\right) +\exp \left( -i\frac{H_{\downarrow }}{\hbar }\tau \right)
\right) \left| \uparrow ,\psi _{i}\right\rangle $. The population
of state $ \left| \uparrow \right\rangle $ is then,
\begin{equation}
P_{\uparrow }=\frac{1}{2}\left( 1+\left| \left\langle \psi
_{i}\left|
e^{i\left( \frac{H_{\downarrow }}{\hbar }\right) \tau }e^{-i\left( \frac{%
H_{\uparrow }}{\hbar }\right) \tau }\right| \psi _{i}\right\rangle
\right| \cos \left( \Delta \tau + \phi_{i} \left( \tau \right)
\right) \right) , \label{ram}
\end{equation}
where $\phi_{i} \left( \tau \right)$ is a phase depending on the
initial state $\left| \psi _{i}\right\rangle $ and the dynamics of
it in the trap. By choosing $\Delta$ sufficiently large, compared
to variations in $\phi_{i} \left( \tau \right)$ Eq. \ref{ram}
yields the well known cosinusodial Ramsey-fringes as a function of
$\Delta \tau $, but with a fringe contrast reduced to the square
root of the fidelity of the external motion in the trap
$P_{FID}\equiv\left| \left\langle \psi _{i}\left| e^{i\left(
\frac{H_{\downarrow }}{\hbar }\right) \tau }e^{-i\left( \frac{
H_{\uparrow }}{\hbar }\right) \tau }\right| \psi _{i}\right\rangle
\right| ^{2}$. In the case where $\left| \psi _{i}\right\rangle $
is an eigenstate of $H_{\downarrow }$, the fidelity can be written
$P_{FID}=\left| \left\langle \psi _{i}\left| e^{-i\left( \frac{
H_{\uparrow }}{\hbar }\right) \tau }\right| \psi _{i}\right\rangle
\right| ^{2}\equiv P_{SRV}$ and is therefore often denoted the
survival probability of $ \left| \psi _{i}\right\rangle $ in
$H_{\uparrow }$. Because the fidelity shows up in many such simple
experiments, especially in connection to quantum information
processing schemes where it measures the overlap between the
desired state $e^{-i\left( \frac{ H_{d }}{\hbar }\right) \tau
}\left| \psi _{i}\right\rangle$ and the actual state $e^{-i\left(
\frac{ H_{a }}{\hbar }\right) \tau }\left| \psi
_{i}\right\rangle$, and because it can be used as an indicator of
quantum chaos, its decay have been an immensely studied topic
throughout \cite{emerson02}.

However, experiments are often not performed with atoms initially
prepared in a pure initial state $\left| \psi _{i}\right\rangle $,
but with atoms in a thermal mixture of states. Since $\phi_{i}
\left( \tau \right)$ depends on the initial state, then the fringe
contrast in a Ramsey experiment will decay rapidly, not due to a
decay in the fidelity of the individual states thermally
populated, but because the cosine terms from different populated
states get out of phase \cite{andersen03}. This rapid loss of
coherence is a problem in quantum information schemes
\cite{kuhr03} and prevents the use of Ramsey spectroscopy
for studying quantum dynamics \cite{andersen03}. The loss in coherence may be overcome by adding an additional potential making $%
H_{\uparrow }=H_{\downarrow }$, and thereby decoupling the
external dynamics from the spectroscopic evolution \cite{ido00}.
This is however not always possible and does not enable the use of
MW-spectroscopy for dynamical studies. The echo pulse sequence
demonstrated in \cite{rowe02}\cite{andersen03}\cite{kuhr03}
provides a more general solution to suppress the inhomogeneous
dephasing of superposition states.

The echo pulse sequence consists of three short pulses ($\frac{\pi
}{2}-\pi-\frac{\pi }{2}$), separated by two dark periods of equal
duration $\tau $, after which the population of $\left| \uparrow
\right\rangle $ is measured. In analogy to Eq. \ref{ram} the
population of state $\left| \uparrow \right\rangle $ can be
written as:
\begin{equation}
P_{\uparrow }=\frac{1}{2}\left[ 1-%
\mathop{\rm Re}%
\left( \left\langle \psi _{i}\left| e^{i\left( \frac{H_{\downarrow }}{\hbar }%
\right) \tau }e^{i\left( \frac{H_{\uparrow }}{\hbar }\right) \tau
}e^{-i\left( \frac{H_{\downarrow }}{\hbar }\right) \tau }e^{-i\left( \frac{%
H_{\uparrow }}{\hbar }\right) \tau }\right| \psi _{i}\right\rangle \right) %
\right]  \label{echo}
\end{equation}
In the following we denote the quantity $F_{ECH}=\left\langle \psi
_{i}\left|
e^{i\left( \frac{H_{\downarrow }}{\hbar }\right) \tau }e^{i\left( \frac{%
H_{\uparrow }}{\hbar }\right) \tau }e^{-i\left(
\frac{H_{\downarrow }}{\hbar }\right) \tau }e^{-i\left(
\frac{H_{\uparrow }}{\hbar }\right) \tau }\right| \psi
_{i}\right\rangle $ the ''echo amplitude''. $F_{ECH}=1$ indicates
complete preservation of coherence and yields $P_{\uparrow}=0$,
whereas $F_{ECH}=0$, yielding $P_{\uparrow}=\frac{1}{2}$ indicates
complete loss of coherence \cite{note2}. From Eq. \ref{echo} it is
seen that
the echo amplitude is insensitive to the addition of constants to $H_{\uparrow }$ and $%
H_{\downarrow }$, so in contrary to the Ramsey signal (Eq.
\ref{ram}) $P_{\uparrow}$ of Eq. \ref{echo} does not depend on
$\omega _{MW}$ and $E_{HF}$, but only on the quantum dynamics of
the atoms in the potentials. The rapid loss of coherence when a
thermal mixtures of states is considered has vanished, making echo
techniques of interest in experiments on quantum information
processing \cite{kuhr03}. This also means that even for thermal
mixtures of states echo spectroscopy can yield information on the
quantum dynamics of the trapped atoms, as was demonstrated in
\cite{andersen03} for more than $10^6$ thermally populated states.

\section{ The Connection Between the Echo and the Survival Amplitudes and long time behavior.}

\label{per}In this section a second order perturbative expression
for the echo amplitude of an eigenstate is derived, showing that
it can be written as a function of the fidelity or survival
amplitude, thereby also predicting the time average of the echo
amplitude.

Since thermal atoms can be considered as a mixture of eigenstates,
the attention is now turned to the echo amplitude of an eigenstate
of $H_{\downarrow }$.
The echo amplitude of an eigenstate $\left| m_{\downarrow }\right\rangle $ of $%
H_{\downarrow }$\ can be rewritten:
\begin{equation}
F_{ECH}\left( \left| m_{\downarrow }\right\rangle ,\tau \right)
=\sum_{l,k,n}\exp \left( i\left( \omega _{m}^{\downarrow }+\omega
_{l}^{\uparrow }-\omega _{k}^{\downarrow }-\omega _{n}^{\uparrow
}\right) \tau \right) \left\langle m_{\downarrow }\mid l_{\uparrow
}\right\rangle \left\langle l_{\uparrow }\mid k_{\downarrow
}\right\rangle \left\langle k_{\downarrow }\mid n_{\uparrow
}\right\rangle \left\langle n_{\uparrow }\mid m_{\downarrow
}\right\rangle  \label{ech2}
\end{equation}
Since adding constants to $H_{\uparrow }$ and $H_{\downarrow }$ does not
change the echo we can without loss of generality set $\omega _{m}^{\uparrow
}=\omega _{m}^{\downarrow }=0$. Now for atoms trapped in far off resonance
optical dipole traps made of linearly polarized light the difference between $%
H_{\uparrow }$ and $H_{\downarrow }$ is very small, so a
perturbative result is of great interest. For a small perturbation
$\left\langle n_{\uparrow }\mid n_{\downarrow }\right\rangle
\simeq 1-\epsilon _{n}^{2}$ and $\left\langle k_{\uparrow }\mid
n_{\downarrow }\right\rangle $ is of the order $\epsilon $ or
less. Rewriting Eq. \ref{ech2} and keeping only terms up to order
$\epsilon ^{2}$ yields:
\begin{eqnarray}
F_{ECH}\left( \left| m_{\downarrow }\right\rangle ,\tau \right)
&\simeq &2\sum_{n\neq m}\exp \left( -i\omega _{n}^{\uparrow }\tau
\right) \left| \left\langle n_{\uparrow }\mid m_{\downarrow
}\right\rangle \right| ^{2}+
\nonumber \\
&&\sum_{n\neq m}\exp \left( -i\omega _{n}^{\downarrow }\tau \right) \left|
\left\langle n_{\downarrow }\mid m_{\uparrow }\right\rangle \right| ^{2}+
\nonumber \\
&&\sum_{n\neq m}\exp \left( i\omega _{n}^{\uparrow }\tau \right) \left|
\left\langle n_{\uparrow }\mid m_{\downarrow }\right\rangle \right| ^{2}-
\nonumber \\
&&\sum_{n\neq m}\exp \left( i\left( \omega _{n}^{\uparrow }-\omega
_{n}^{\downarrow }\right) \tau \right) \left| \left\langle n_{\uparrow }\mid
m_{\downarrow }\right\rangle \right| ^{2}-  \nonumber \\
&&\sum_{n\neq m}\exp \left( -i\left( \omega _{n}^{\uparrow }+\omega
_{n}^{\downarrow }\right) \tau \right) \left| \left\langle n_{\uparrow }\mid
m_{\downarrow }\right\rangle \right| ^{2}+  \nonumber \\
&&\left| \left\langle m_{\uparrow }\mid m_{\downarrow }\right\rangle \right|
^{4}  \label{longex}
\end{eqnarray}
This rewriting does not look like a simplification of Eq.
\ref{ech2}, but first it is noted that the echo is no longer a
function of $\ $matrix elements like $\left\langle n_{\uparrow
}\mid m_{\downarrow }\right\rangle $ but only a of their absolute
value square, often referred to as the local density of states
(LDOS) \cite{Cohen01}. Second, note that since $\omega
_{m}^{\uparrow }=\omega _{m}^{\downarrow }=0$, then typically
$\omega _{n}^{\uparrow }-\omega _{n}^{\downarrow }<<\omega
_{n}^{\downarrow }$, so except for extreme long times the
approximation $\omega _{n}^{\uparrow }\simeq \omega
_{n}^{\downarrow }\equiv \omega _{n}$ can be used. For high
excited states $\sum_{n\neq m}\exp \left( i\left( \omega
_{n}^{\uparrow }\right)\tau \right) \left| \left\langle
n_{\uparrow }\mid m_{\downarrow }\right\rangle \right|
^{2}=\sum_{n\neq m}\exp \left( -i\left( \omega _{n}^{\uparrow
}\right) \tau \right) \left| \left\langle n_{\uparrow }\mid
m_{\downarrow }\right\rangle \right| ^{2}$ is also a good
approximation, for times much shorter than $1/\Delta _{mls}$ with
$\Delta _{mls}$ the mean level spacing. Using these approximations
Eq. \ref{longex} can be rewritten as:
\begin{eqnarray}
F_{ECH}\left( \left| m_{\downarrow }\right\rangle ,\tau \right)
&\simeq &4\sum_{n\neq m}\exp \left( -i\omega _{n}\tau \right)
\left| \left\langle
n_{\uparrow }\mid m_{\downarrow }\right\rangle \right| ^{2}-  \nonumber \\
&&\sum_{n\neq m}\exp \left( -2i\omega _{n}\tau \right) \left| \left\langle
n_{\uparrow }\mid m_{\downarrow }\right\rangle \right| ^{2}+  \nonumber \\
&&\left| \left\langle m_{\uparrow }\mid m_{\downarrow
}\right\rangle \right| ^{6},  \label{shex}
\end{eqnarray}
where $\left| \left\langle m_{\uparrow }\mid m_{\downarrow
}\right\rangle \right|
^{6}\simeq 1- 6 \epsilon _{m} ^2$. The quantity $\left\langle m_{\downarrow }\left| \exp \left( -i\left( \frac{%
H_{\uparrow }}{\hbar }- \omega _{m}^{\uparrow } \right) \tau
\right) \right| m_{\downarrow }\right\rangle $ is denoted the
survival amplitude ($F_{SRV}$), since its length square is the
survival probability. Writing the survival amplitude in the same
way as Eq. \ref{shex} shows that Eq. \ref{shex} yields a simple
relation between the echo amplitude and the survival amplitude of
an eigenstate in the perturbative limit:
\begin{eqnarray}
F_{ECH}\left( \left| m_{\downarrow }\right\rangle ,\tau \right)
&\simeq& 4F_{SRV}\left( \left| m_{\downarrow }\right\rangle ,\tau
\right)-F_{SRV}\left( \left| m_{\downarrow }\right\rangle ,2\tau
\right)-2 \label{etsp}
\end{eqnarray}
Eq. \ref{etsp} indicates that the time average of the echo
amplitude is always smaller than the time average of the survival
amplitude. Hence in the low temperature regime, where the decay of
coherence is set by the decay of the survival amplitude, and not
by the thermal spread of initial conditions, a Ramsey type
experiment will yield better coherence than an echo experiment
(see Eq. \ref{ram}). This is in agreement with the experimental
results published in \cite{rowe02}.

From Eq. \ref{shex} it is seen that for times $\tau $ where the
approximation $\omega _{n}^{\uparrow }\simeq \omega
_{n}^{\downarrow }$ is valid, the constant contribution to the
echo signal is not given by the infinite long time limit $\left|
\left\langle m_{\uparrow }\mid m_{\downarrow }\right\rangle
\right| ^{4}$ but by $\left| \left\langle m_{\uparrow }\mid
m_{\downarrow }\right\rangle \right| ^{6}$. This is a rather
peculiar result since $\left| \left\langle m_{\uparrow }\mid
m_{\downarrow }\right\rangle \right| ^{6}<\left| \left\langle
m_{\uparrow }\mid m_{\downarrow }\right\rangle \right| ^{4}$ hence
it implies that waiting for dephasing to occur on time scales of
$1/\left( \omega _{n}^{\uparrow }-\omega _{n}^{\downarrow }\right)
$ will increase the ensemble average of the echo coherence. For
perturbations that conserve the eigenspectrum the $\left|
\left\langle m_{\uparrow }\mid m_{\downarrow }\right\rangle
\right| ^{6}$ echo level persists for all times.

In \cite{andersen03} we compared the measured long time echo
signal for atoms confined in a Gaussian laser beam to a
calculation based on the ensemble average of the infinite time
level of $\left| \left\langle m_{\uparrow }\mid m_{\downarrow
}\right\rangle \right| ^{4}$. However, approximating the Gaussian
trap of \cite{andersen03} by a harmonic oscillator yields that
typically $2\pi / \left( \omega _{n}^{\uparrow }-\omega
_{n}^{\downarrow }\right) \sim 5s$, whereas the long time echo
level was measured when the initial oscillations had died out,
typically after $\tau \sim 10 ms$. Hence, from the perturbative
treatment we expect to observe the ensemble averaged $\left|
\left\langle m_{\uparrow }\mid m_{\downarrow }\right\rangle
\right| ^{6}$ echo level rather than the $\left| \left\langle
m_{\uparrow }\mid m_{\downarrow }\right\rangle \right| ^{4}$
level. We repeated the echo-level calculation of \cite{andersen03}
using now the assumption that the long time echo signal is given
by $\left| \left\langle m_{\uparrow }\mid m_{\downarrow
}\right\rangle \right| ^{6}$, and found better agreement with the
experimental data published in \cite{andersen03} as compared to
the $\left| \left\langle m_{\uparrow }\mid m_{\downarrow
}\right\rangle \right| ^{4}$ calculation, but due to the
uncertainty of the data points no definitive conclusion can be
made \cite{note1}.

Finally, the simplified model presented here predicts that if
$\left| \left\langle m_{\uparrow }\mid m_{\downarrow
}\right\rangle \right| ^{6} \simeq 1$ for all occupied states,
then perfect echo coherence will persist for infinite times. In
real systems, there are effects, such as photon scattering or
temporal fluctuations in the trap depth not taken into account in
our model (Eq. \ref{ham}) causing echo coherence to be lost on
long time scales \cite{andersen03} \cite{kuhr03}.

\section{Equation 7, and Decay and Revivals of Echo and Survival Probability.}

\label{fire}In this section the validity of Eq. \ref{etsp} and the
behavior of the quantities involved in it, are investigated by
numerical calculations on a chaotic hard wall Burnimovich stadium
\cite{burni79}. This system is used as a test system, since the
Burnimovich stadium has become almost a synonym with classical
chaotic billiard dynamics, and since the scaling method of
\cite{vergini95}\cite{barnett00} enables efficient computation of
high excited eigenstates of this system. It is demonstrated that
the echo amplitude and the survival probability show partial
revivals at certain times for a spacial shift perturbation.

\begin{figure}[tbp]
\includegraphics[width=3inc]{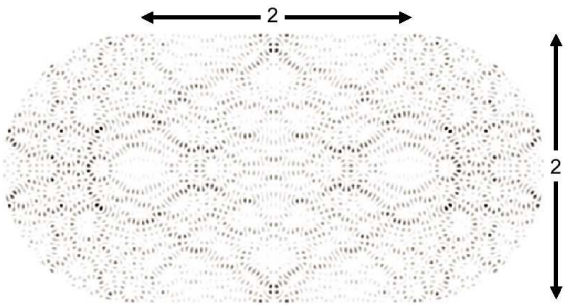}
\caption{Highly excited eigenstate in a Burnimovich billiard
(excitation number $\sim11000$).} \label{fi3}
\end{figure}

The stadium considered consists of two semicircles of radius 1,
connected by two lines of length 2 (see fig. \ref{fi3}). We
consider the perturbation caused by shifting $V_{\uparrow }\left(
{\bf x}\right) $ a small amount along the long axis of the stadium
compared to $V_{\downarrow }\left( {\bf x}\right) $. This
perturbation conserves the eigenspectrum so $\omega _{n}^{\uparrow
}=\omega _{n}^{\downarrow }$. The shift also has direct relevance
to the ''dephasing free'' atom optics billiard proposed in \cite
{andersen03b}, and illustrates some of the points of later
sections.

The computation of the echo and the survival probability is done
in the straight forward way. A large cluster containing $\sim
1300$ eigenstates with energies around 11000 is computed using the
scaling method of \cite{vergini95}. The units
h\hskip-.2em\llap{\protect\rule[1.1ex]{.325em}{.1ex}}\hskip.2em
=2m=1 are used so the eigenstates in the interior of the billiard
are solutions to $\nabla ^{2}\left| \psi \right\rangle
=-k^{2}\left| \psi \right\rangle $ and are vanishing on the
boundary. The transformation matrix $T$ taking states from a
representation in eigenstates of \ $H_{\uparrow }$ to a
representation in eigenstates of $H_{\downarrow }$ is then
computed by computing the overlap integral of the eigenstates of
$H_{\uparrow }$ with those of $H_{\downarrow }$ (the length square
of the elements of $T$ being the LDOS \cite{Cohen01}). The
computation of the overlaps is efficiently done using the methods
of \cite{barnett00}. Since the time evolution of a state in the
representation of eigenstates of the Hamiltonian under which it
evolves is trivial, it is a trivial task to compute the echo and
the survival amplitude when $T$ is known using Eq. \ref{etsp}. The
issue of $T$ not being completely unitary for a hard wall billiard
is addressed in \cite{Cohen01}.

\begin{figure}[tbp]
\includegraphics[width=3inc]{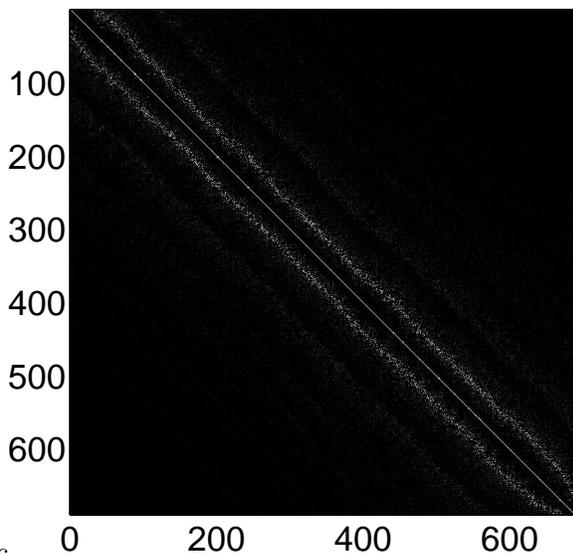}
\caption{$|\left\langle k_{\uparrow }\mid n_{\downarrow
}\right\rangle|^2$ (the elements of the transform matrix T squared
(see text)) for 700 states centered around $n=m\sim11000$ for the
Bunimovich stadium of Fig. 2, perturbed by a shift of 0.004.
Distinct sidebands are observed.} \label{fi4}
\end{figure}

\begin{figure}[tbp]
\includegraphics[width=3inc]{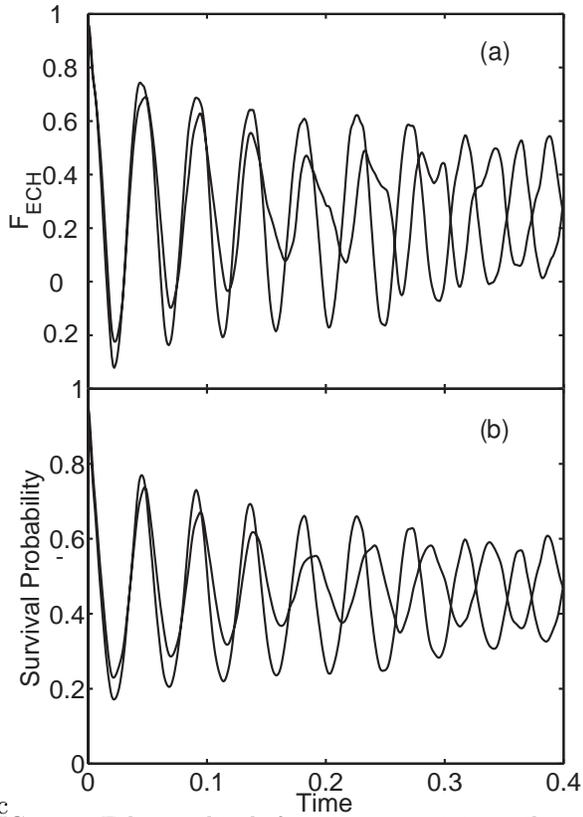}
\caption{a: Echo amplitude from two eigenstates with similar
energies in a Burnimovich billiard (excitation number
$\sim11000$), and the perturbation being a shift of 0.008. Clear
revivals are seen. b: The same behavior is seen for the survival
probability as expected from Eq. 7.} \label{fi5}
\end{figure}

The structure of $T$ can depend on the perturbation even when the
underlying classical dynamics is chaotic
\cite{Cohen01}\cite{barnett00} \cite{Barnett00b}. In Fig.
\ref{fi4} a plot of the square of the elements of $T$ is shown for
the perturbation of a shift of 0.004, where two distinct sidebands
are seen. This structure induces special
behavior of the echo and the survival probability, since an eigenstate of $%
H_{\downarrow }$ will show a (partial) revival when evolved in $H_{\uparrow
} $ at the time the sidebands have acquired a phase of 2$\pi $. This is
demonstrated in Fig \ref{fi5}, where the echo and survival probability for
several different eigenstates are shown. They all show revivals at the time $%
\tau _{R} $\ corresponding to the sidebands acquiring a phase of
2$\pi .$ At longer times the revivals from different states get
out of phase due to the finite width of the sidebands in Fig.
\ref{fi4}. We note that the revivals seen in Fig. \ref{fi5} are
not the trivial revivals originating from the system
being discrete, but they appears since the nature of the difference between $%
H_{\downarrow }$ and $H_{\uparrow }$ induces "selection rules"
that determines to which states in $H_{\uparrow }$ an eigenstate
of $H_{\downarrow }$ is coupled.

\begin{figure}[tbp]
\includegraphics[width=6inc]{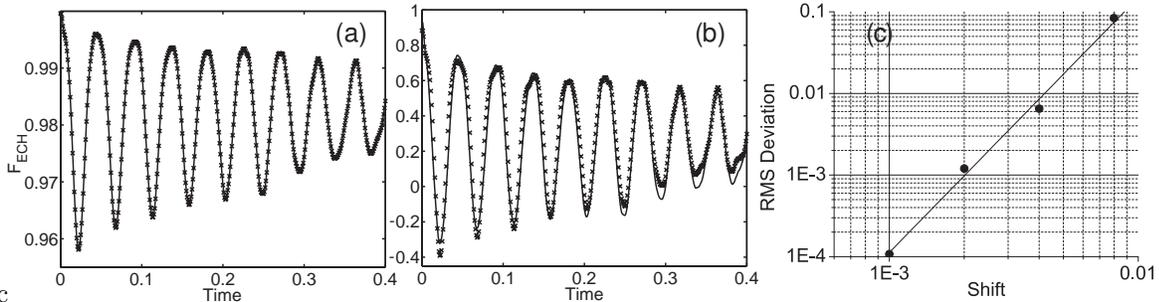}
\caption{a: Solid line: Echo amplitude calculated by full
calculation for a shift of 0.001 of a Bunimovich stadium. x:
approximation by Eq. 7. Excellent agreement is seen. b: The same
as (a) but for a shift of 0.008. Good agreement still exist in
spite of very strong oscillations. c: RMS deviation between exact
Echo amplitude and perturbative approximation for time 0-0.1 as a
function of perturbation (x). It is seen to agree with
$\protect\epsilon^{3}$ behavior (solid line).} \label{fi6}
\end{figure}

Shown in Fig. \ref{fi6} is the perturbative approximation Eq.
\ref{etsp} to the echo, together with the exact calculations for a
shift of 0.001 (Fig. \ref{fi6}a) and 0.008 (Fig. \ref{fi6}b),
showing good agreement, even for large perturbations. In
\ref{fi6}c the RMS deviation between the echo and the
approximation  (Eq. \ref{etsp}) for the time interval 0-0.1 is
shown as a function of the shift. It is seen that the deviation
grows as $\epsilon ^{3}$, as expected from Eq. \ref{etsp} being a
second order approximation to the echo. This indicates that the
other approximations made are valid.

\begin{figure}[tbp]
\includegraphics[width=3inc]{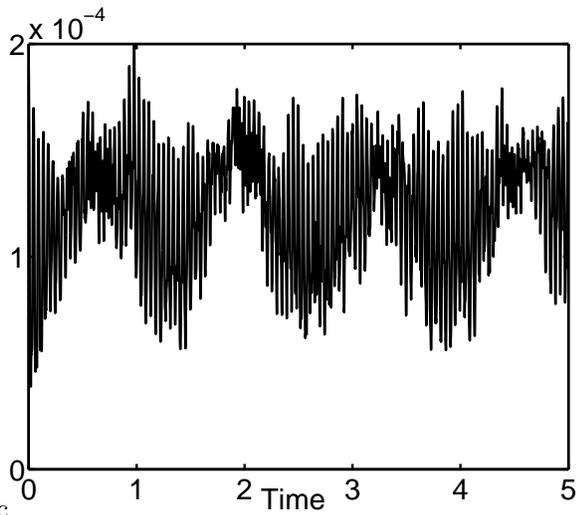}
\caption{The difference between echo amplitude and its
perturbative approximation as a function of time for a shift of
0.001 of the Bunimovich Stadium. No divergence is observed even
for long times.} \label{fi7}
\end{figure}

In Fig. \ref{fi7} the difference between the exact echo and the
approximation (Eq. \ref{etsp}) is shown, for long times, for one
of the states of Fig. \ref{fi5}. No divergence as a function of
time is observed. This is subscribed to the fact that the terms
neglected in order to obtain Eq. \ref{etsp} are oscillatory in
time.

\section{Semi-Classical Estimation of the Echo Amplitude}

\label{fem}In this section the echo amplitude of an ensemble of
high excited eigenstates is calculated using classical estimations
of the LDOS \cite{Cohen01}. The method is confirmed by comparison
to a full quantum calculation on a hard-wall Burnimovich stadium.
It is then used in section \ref{seks} to calculate the echo in a
chaotic atom optics billiard, that have been demonstrated
experimentally in the past \cite{milner01}\cite{Davidson95}.
Revivals in analogy with those of the previous section are found
to appear.

From the fact that the perturbative limit is considered so $\overline{\left|
\left\langle m_{\uparrow }\mid m_{\downarrow }\right\rangle \right| ^{6}}%
\simeq \overline{\left| \left\langle m_{\uparrow }\mid
m_{\downarrow }\right\rangle \right|^2 }^{3}$ (where
$\overline{x}$ stands for the average of $x$) it is apparent that
Eq.s \ref{longex}-\ref{etsp} also hold for averages. This means
that classical methods
\cite{Cohen01}\cite{barnett00}\cite{Barnett00b} can be used to
estimate the average LDOS and from this the averaged echo
amplitude can be computed.

A detailed description of the computation of the classical
estimations of the LDOS is given in
\cite{Cohen01}\cite{barnett00}\cite{Barnett00b}, so we only give a
brief review of the steps involved. Writing $H_{\uparrow
}=H_{\downarrow }+\epsilon H_{p}$ then first order perturbation
theory gives that:
\begin{equation}
\left| \left\langle n_{\uparrow }\mid m_{\downarrow }\right\rangle
\right| ^{2}\simeq \frac{\epsilon ^{2}}{\Delta E_{n,m}^{2}}\left|
\left\langle n_{\downarrow }\right| H_{p}\left| m_{\downarrow
}\right\rangle \right| ^{2} \label{perr}
\end{equation}
For high excited states the band profile of the matrix $\left|
\left\langle n_{\downarrow }\right| H_{p}\left| m_{\downarrow
}\right\rangle \right| ^{2}$ can be estimated classically as:
\begin{equation}
\overline{\left| \left\langle n_{\downarrow }\right| H_{p}\left|
m_{\downarrow }\right\rangle \right| ^{2}} \left( \Delta
E_{n,m}\right) \simeq \frac{\Delta _{mls}}{2\pi \hbar
}\widetilde{C}\left( \frac{\Delta E_{n,m}}{\hbar }\right),
\label{dor}
\end{equation}
where the average is taken over few adjacent states, $\Delta
_{mls}$ is the mean level spacing and $\widetilde{C}\left( \omega
\right) $ is determined as following. A long classical trajectory
${\bf x}\left( t\right) $ is computed, and $-H_{p}$ is evaluated
along it. The time average $F_{m}$ is subtracted to yield the
fluctuating quantity:
\begin{equation}
F\left( t\right) =-H_{p}\left( {\bf x}\left( t\right) \right) -F_{m}
\end{equation}
The autocorrelation of $F\left( t\right) $ is called $C\left(
t\right) $, and $\widetilde{C}\left( \omega \right) $, its Fourier
transform, can be computed as the power spectrum of $F\left(
t\right) $. $\widetilde{C}\left( \omega \right) $ is an ensemble
average of all classical trajectories of energy $E_{m}$, but since
a chaotic system is considered, this can be taken as one long
trajectory. The deeper reasons for the validity of Eq. \ref{dor},
and its good accuracy, have been studied in detail \cite{Cohen01}
\cite{barnett00}\cite{Barnett00b}. Combining Eq. \ref{dor} and
\ref{perr} yields the expression:
\begin{equation}
\overline{\left| \left\langle n_{\uparrow }\mid m_{\downarrow
}\right\rangle \right| ^{2}}\left( \Delta E_{n,m}\right) \simeq
\frac{\epsilon ^{2}}{\Delta
E_{n,m}^{2}}\frac{\Delta _{mls}}{2\pi \hbar }\widetilde{C}\left( \frac{%
\Delta E_{n,m}}{\hbar }\right)  \label{meo}
\end{equation}
$\overline{\left| \left\langle m_{\uparrow }\mid m_{\downarrow
}\right\rangle \right| ^{2}}$ is then found as:
\begin{equation}
\overline{\left| \left\langle m_{\uparrow }\mid m_{\downarrow }\right\rangle
\right| ^{2}}\simeq 1-\frac{\epsilon ^{2}}{2\pi \hbar ^{2}}%
\mathrel{\mathop{\int }\limits_{\left| \omega \right| >\frac{\Delta _{mls}}{\hbar }}}%
\widetilde{C}\left( \omega \right) \omega ^{-2}d\omega  \label{mm}
\end{equation}

\begin{figure}[tbp]
\includegraphics[width=3inc]{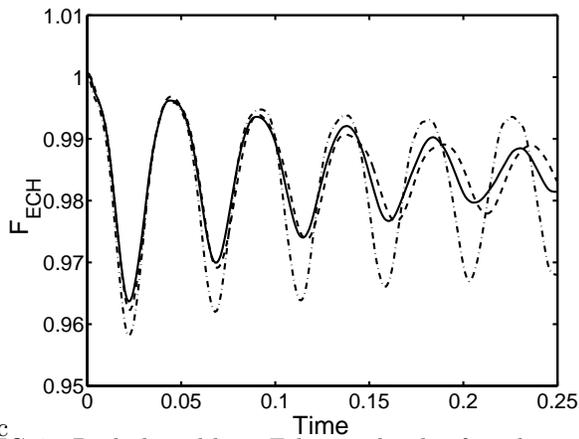}
\caption{Dash-dotted line: Echo amplitude of single eigenstate for
a Bunimovich stadium and a shift of 0.001. Dashed line:
Semiclassical approximation using Eq. 11 and 12 in Eq. 6. Solid
line: Ensemble average of echo amplitudes from 15 adjacent
eigenstates. The semiclassical approximation does not resemble the
echo from a single eigenstate, but closely resembles the average
of just 15 eigenstates. No fitting or scaling is used.}
\label{fi8}
\end{figure}

Using Eq. \ref{meo} and \ref{mm} in Eq. \ref{shex} or \ref{etsp},
then allows for calculation of the echo amplitude averaged over a
narrow energy band. A comparison between a semiclassical
calculation of the echo amplitude using Eq. \ref{meo} and \ref{mm}
to the full calculation of the previous section for the shift
perturbation of a hard-wall Bunimovich billiard, is shown in fig
\ref{fi8}. It is seen that the semiclassical result does not
resemble the echo of an individual eigenstate, but shows good
agreement with an average of 15 adjacent eigenstates. Since
''hot'' atoms are considered the population of eigenstates is a
slowly varying function of the eigenstate number, thereby enabling
the ensemble-averaged echo signal to be computed. The evaluation
of $\widetilde{C}\left( \omega \right) $ for the hard wall
billiard is done following Refs.
\cite{Cohen01}\cite{barnett00}\cite {Barnett00b}. The revivals
occur at times $\sim 2\tau _{bl}$ with $\tau _{bl}$ being the
ballistic time.

The shift conserves the eigenspectrum of the billiard, hence
mixing to nearby states is suppressed \cite{andersen03b}. For a
more generic perturbation such mixing effects can
occur, thereby drastically reducing the range of perturbation where the approximation $%
\left\langle m_{\uparrow }\mid m_{\downarrow }\right\rangle \simeq
1-\epsilon _{m}^{2}$ is valid \cite{Cohen01}\cite{andersen03b}.
However, the major part of the probability will still be contained
in a very narrow ''core'' and for short timescales this core will
not have time to dephase. Therefore we can neglect the decay of
the core and replace $\left| \left\langle m_{\uparrow
}\mid m_{\downarrow }\right\rangle \right| ^{6}$ in Eq. \ref{shex} with $%
\left| \left\langle \psi _{\uparrow }^{c}\mid m_{\downarrow
}\right\rangle \right| ^{6}$ where $\left| \psi _{\uparrow
}^{c}\right\rangle $ is the core state, sum only over states
outside the core, and still get a reasonable estimation of the
echo for short times.

\section{The Light-Sheet Wedge Billiard}

\label{seks}
\begin{figure}[tbp]
\includegraphics[width=3inc]{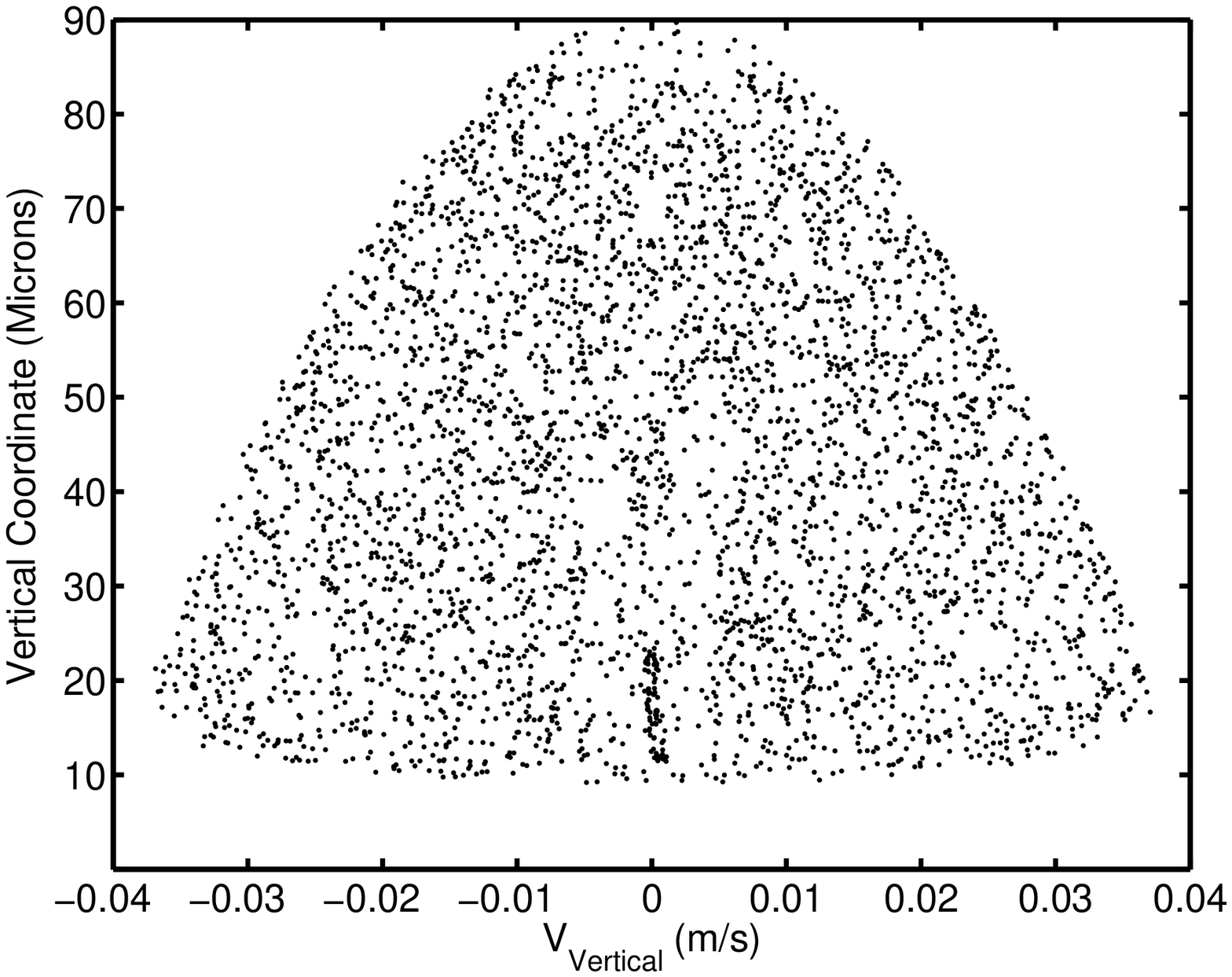}
\caption{Poincar\'{e} surface of section for the light-sheet wedge
billiard described in the text. The vertical position is plotted
as a function of the vertical momentum, when the atom crosses the
symmetry axis of the wedge, for a single long trajectory. Phase
space is seen to be chaotic.} \label{fi9}
\end{figure}

Now the attention is turned to an experimental
atom-optics billiard, namely the light-sheet wedge
\cite{milner01}\cite{Davidson95}. We consider a wedge made from
two elliptical Gaussian beams with widths of 10 $\mu m$ by 150
$\mu m$. The wedge points downwards so gravity confines atoms in
it. The wedge angle is 98 degree and the sheets cross 67 $\mu m$
from the centers. Only two dimensions are considered, so the
ballistic time in the third dimension is assumed to be much longer
than the experiment time (as in Ref. \cite{andersen03}).
Alternatively, the atoms can be confined in the third dimension by
a very far off resonance one dimensional optical lattice, that
essentially freezes the motion in that direction
\cite{Friedman01}. A long classical trajectory is computed for a
$^{85}Rb$ atom with an energy of 1.1$\times 10^{-28}$ J thus
reaching a maximal height of 80 $\mu m$ above the bottom of the
wedge. The laser power of the wedge is taken to be 10 mW and the
trap laser wavelength 779 nm (the resonant wavelength of the
$D_{2}$-line is 780 nm). In fig \ref{fi9} a Poincar\'{e} surface
of section is presented showing that the classical motion of the
atom in this billiard is indeed chaotic in spite of it not being
the idealized wedge billiard \cite {Lehtihet86}.

The two states considered $\left| F=2,m_{F}=0\right\rangle $
($\left| \downarrow \right\rangle $) and $\left| F=3,m_{F}=0
\right\rangle $ ($\left| \uparrow \right\rangle $)
are separated by the hyperfine splitting of 3 GHz. Due to the difference in strength of dipole potential $H_{\uparrow }$ and $%
H_{\downarrow }$ are related $H_{\uparrow }=H_{\downarrow
}+\epsilon U_{dipole} $ with $\epsilon=6\times10^{-3}$ for the
parameters considered.

If the wedge had been made from exponentially decaying evanescent
waves close to a prism surface instead of Gaussian light-sheets
then the change in strength in dipole potential can be described
as a shift of the entire potential along the vertical axis. Since
the shift perturbation conserves the eigenspectrum such traps are
expected to yield very long coherence times and the use of echo
techniques would not be necessary \cite{andersen03b}. However, for
a trap made from Gaussian light-sheets as we here consider, the
perturbation does not have a simple geometric interpretation and
does not conserve the eigenspectrum. Therefore echo techniques is
expected to be of relevance, and the effects demonstrated is not
expected to be limited to eigenspectrum conserving perturbations.

\begin{figure}[tbp]
\includegraphics[width=3inc]{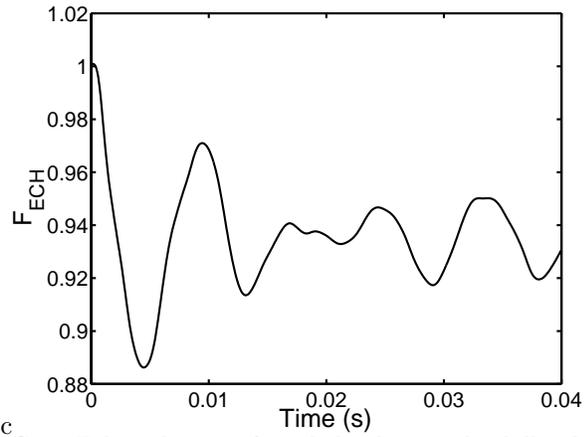}
\caption{Echo coherence for a light sheet wedge billiard described
in the text. In spite of the underlying classical dynamics being
chaotic oscillations are seen.} \label{fi10}
\end{figure}

Using the classical trajectory the echo amplitude is predicted by
the methods of the previous section. The results are shown in Fig.
\ref {fi10} ($\left| \left\langle m_{\uparrow }\mid m_{\downarrow
}\right\rangle \right| ^{6}\simeq 0.96$). In analogy to the
hard-wall billiard example given above, the echo is seen to revive
at times around the ballistic time. This is the same phenomena as
was observed in \cite{andersen03} for atoms in a Gaussian trap.
These results indicates that the revivals demonstrated in
\cite{andersen03}, are not due to the Gaussian trap being nearly
harmonic, but a more generic feature of dipole traps even when the
underlying dynamics is chaotic.

\section{Conclusions}

A perturbative expression for the micro wave echo coherence for
optically trapped atoms was derived, and confirmed numerically,
showing that the echo signal can be written as a function of the
survival probability or fidelity of eigenstates of the motion of
the atoms in the trap. Based on this perturbative result the long
time echo coherence of atomic ensembles was predicted. It was
demonstrated that in the perturbative limit the echo signal and
the survival probability display ''system specific'' features,
despite the fact that the underlying classical dynamics is
chaotic. This was done numerically by demonstrating partial
revivals in the echo signal and the survival probability for a
perturbation consisting of a small shift of the potential. Based
on the perturbative expression for the echo signal a
semi-classical method for calculating the averaged echo signal was
presented. The validity of this expression was confirmed
numerically for a hard wall Burnimovich billiard. The method was
then used to calculate the echo signal for atoms in an
experimentally realizable atom-optics billiard, namely the
light-sheet wedge billiard. Revivals in the echo coherence were
found to appear in this system, leading to the conclusion that the
revivals demonstrated in \cite{andersen03}, are not limited to
nearly harmonic traps, but are a generic feature of dipole traps.

We gratefully acknowledge helpful discussions and comments from
Uzy Smilansky, Klaus M{\o}lmer and Doron Cohen. This work was
supported by Minerva Foundation, and Foundation Antorchas.

\begin{references}
\bibitem{note1} It is noted that for some of the data points of
\cite{andersen03} the perturbation exceeds strengths where a
second order perturbative result it expected to be valid.
\bibitem{note2} The echo amplitude also characterizes the contrast
of Ramsey-like fringes that occur around $\tau_2=\tau_1$ when the
two dark times are not equal.
\end{references}
\end{document}